\documentclass[]{article}

\usepackage{amsthm,amsmath,natbib}
\usepackage{graphicx}
\usepackage{dcolumn}
\usepackage{bm}
\usepackage{natbib}
\usepackage[utf8]{inputenc}
\usepackage[T1]{fontenc}
\usepackage{mathptmx}
\usepackage{amssymb}

\begin{document}

\title{An explanation of the Bell experiment.}

\author{Inge S. Helland, \\ Professor emeritus, Department of Mathematics, University of Oslo}

\date{}

\maketitle

\begin{abstract}
The Bell experiment is explained in the light of a new approach to the foundation of quantum mechanics. It is concluded from my basic model that the mind of any observer must be limited in some way: In certain contexts, he is simply not able to keep enough variables in his mind when making decisions. This has consequences for Bell's theorem, but it also seems to have wider consequences.

\end{abstract}


\section{Introduction}

A completely new approch towards the foundation of quantum theory was proposed in Helland (2023b). Part of the mathematical results of that article can help us to understand the surprising results of the famous Bell experiment.

The Nobel Prize in Physics for 2022 was given for performing different versions of the Bell experiment, and this has caused much attention among both laymen and scientists. This experiment was done with entangled quantum states, where two particles belong together as a unit also when they are separated. The experiments have shown without any doubt that what are called Bell's inequalities may be violated in practice. These inequalities may be derived classically by seemingly obvious arguments.

Despite an enormous literature on the subject, a real understanding of the result of the experiment is still lacking. The purpose of this article is to propose a solution, not by being a Bell denier, and not by assuming some action at a distance, but simply by departing from a general realist view of the world. In my opinion, every description of a physical phenomenon should be from the point of view of an observer or from the point of view of a set of communicating observers. This is in agreement with the philosophy of Convivial Solipsism, introduced by Zwirn (2016) as an attempt to solve the measurement problem and related problems.

My own background is as a statistician, but I have now worked for many years on problems related to the understanding of quantum mechanics. One of my aims has been to find some basis for quantum theory that I am able to explain to my fellows, the statisticians. As a step towards this, I published on Springer in 2018 the book 'Epistemic Processes'. Later I discovered that Chapter 4 in that book was incomplete and contained some serious errors. These errors have now been corrected in a revised 2021 edition of the book. In the last two years I have published a number of articles related to my views on quantum foundations; some of these articles are collected in the book Helland (2023a). All my arguments together with some new results that simplify my theory considerably are now gathered in the new article Helland (2023b).
 
Central notions for me are the questions that we ask nature.
 The whole of empirical science is in reality the field of asking questions to nature and trying to obtain answers. The purpose of statistics is to help empirical science in this process. In some way, I feel that there should be a connection between theoretical sciences whose purpose is related to the study of nature on the one hand and statistical theory on the other hand. This is one point of departure from my books and papers.
 
To me, it is also natural that an ultimate quantum theory should be related to asking questions about nature. The first problem raised in my technical note Helland (2019) is whether this can be seen to be the case with the existing quantum theory. And the answer to this seems largely to be affirmative. For spin components in the spin 1/2 case, the one-to-one correspondence between quantum states on the one hand and question-and-answer pairs on the other hand seems to be quite clear. This is proved in Helland (2021, 2023a) and more generally in H\"{o}hn and Wever (2017).
 
This is the way I try to approach quantum theory in my books and in my papers: Through variables that are related to questions that we can ask about nature. The basic notion is that of theoretical variables, variables contained in this connection in the mind of an observer or in the joint minds of a set of communicating observers. 

The theoretical variables may be accessible or inaccessible. A variable $\theta$ is said to be accessible to an observer $C$ if it in principle is possible for this observer to find as accurate values for $\theta$ as he wants to. A similar definition holds for a communicating set of observers.

In a spin 1/2 situation, for instance, an accessible theoretical variable is nothing but the spin component in a given direction as perceived by an observer (or by communicating observers). However, the full spin vector is inaccessible to any observer.
 
Having a purely epistemic approach towards quantum mechanics, it seems possible to understand so-called paradoxes like Schr\"{o}dinger's cat, Wigner's friend and the two-slit experiment. In Helland (2023a,c) I have given my own solutions, based on a version of quantum mechanics where the only allowable pure state vectors are those that are eigenvectors of some physically meaningful operator.
 
In my approach to the Born formula (Helland, 2021) I assume that the actor in question has an authority that can be said to be perfectly rational. Understanding this more properly seems to be of interest, but this problem will not be discussed here.
 
In my view, seeking a theory that covers reality in all situations - whatever that means, seems to be impossible. We are not gods. Instead, in my opinion, we should be looking for a theory that is related to our knowledge of reality. Any process of seeking knowledge can be called an epistemic process. In my books and articles I have concentrated on epistemic processes connected to measurements or experiments.
 
Quantum theory should be seen as a mathematical model, a model related to the mind of an observer, or to the joint minds of a communicating set of observers.
 
Consider the Bell experiment situation, which will be further explained below. To an observer Charlie, receiving data from both the primitive actors Alice and Bob, there are two accessible theoretical variables: The measured spin component as perceived by Alice, and the measured spin component as perceived by Bob. Alice and Bob cannot communicate during the experiment. Alice has  free will to choose the direction in which she wants to measure the spin component, and so has Bob (or they choose their directions by some random choice). Then they achieve their responses. Quantum theory is related to these perceived responses. Bell's inequalities have to do with some hypothetical reality behind these measurements, and this hypothetical reality, in my opinion, is not completely accessible for any human. As I see it, this is in some way all there is to it.

This article can be seen as an improved version of Helland (2022a). It is closely related to the article Helland (2023d).

\section{On entanglement and EPR}

Consider two spin 1/2 particles, originally in the state of total spin 0, then separated, one particle sent to the actor Alice and one particle sent to the actor Bob. This can be described by the entangled singlet state
\begin{equation}
|\psi\rangle = \frac{|1+\rangle |2-\rangle - |1-\rangle |2+\rangle}{\sqrt{2}},
\label{spin1}
\end{equation}
where $|1+\rangle$ means that particle 1 has spin component +1 in some fixed direction, and $|1-\rangle$ means that the component is -1; similarly, for $|2+\rangle$ and $|2-\rangle$ with particle 2. 

As in David Bohm's version of the EPR situation, let Alice measure the spin component of her particle in some direction $a$, and let Bob measure the spin component of his particle in the same direction. As has been described in numerous papers, there seemingly is a strange correlation here: The spin components are always opposite. 

This can be easily understood if only one direction can be assumed to be measured: It is then like an experiment with two boxes, where, unknown to Alice and Bob, the label `-1' is placed in one box and the label `+1' is placed in the other box. Later the boxes are separated, one box is opened by Alice and the other one by Bob.

What is strange about the EPR experiment, is that opposite spins are measured for absolutely all choices of direction.

I want to couple this with the philosophy of Convivial Solipsism [30]: Every description of the world must be relative to some observer. So let us introduce an observer, Charlie, observing the results of both Alice and Bob. Charlie's observations are all connected to the entangled state (\ref{spin1}). 

In Helland (2023b) the bases for understanding are the notions of accessible and inaccessible theoretical variables. An example is a spin $1/2$ particle, where I base my model on an inaccessible unit spin vector $\phi$, defined such that the spin component in any direction $a$ is the discretized projection of $\phi$ upon $a$.

Let us try to describe the EPR experiment in terms of such accessible and inaccessible variables. The unit spin vectors $\phi_1$ and $\phi_2$ of the two particles are certainly inaccessible to Charlie, but it turns out that their dot product $\eta=\phi_1\cdot \phi_2$ is accessible to him. In fact, Charlie's observations are forced to be related to the state given by $\eta=-3$.

Mathematically this is proved as follows. The eigenvalues of the operator $A^\eta$ corresponding to $\eta$ are 1 and -3. The eigenvector associated with the eigenvalue -3 is just $|\psi\rangle$ of (\ref{spin1}), while the eigenspace associated with the eigenvalue 1 is three-dimensional. (See for instance exercise 6.9. page 181 in Susskind and Friedman, 2014.)

What does it mean that $\eta=\phi_1\cdot \phi_2=-3$? It means that $\phi_{1x}\phi_{2x}+\phi_{1y}\phi_{2y}+\phi_{1z}\phi_{2z}=-3$, and since all components here are either -1 or +1, this is only possible if $\phi_{1x}\phi_{2x}=-1$ etc., which implies $\phi_{1x}=-\phi_{2x}$, $\phi_{1y}=-\phi_{2y}$, and $\phi_{1z}=-\phi_{2z}$. It follows that $\phi_{1a}=-\phi_{2a}$ in every direction $a$. It is assumed here that, even though the derivation is concerned with inaccessible variables, the conclusion, which is an accessible conclusion for some experiment, is nevertheless valid.

Note that Charlie can be any person. So, we conclude: To any observing person, the spin components as measured by Alice and Bob must be opposite. This seems to be a necessary conclusion, implied by the fact that the person, relative to his observations, is related to the state given by (\ref{spin1}).

\section{The Bell experiment}

The two actors Alice and Bob are spacelikely separated at the moment when they observe. Midways between them is a source of entangled spin 1/2 particles, one particle in a pair is sent towards Alice, the other towards Bob. In concrete terms, the joint state of the two particles is given by (\ref{spin1}).

Here and in the following the spin component in any direction is normalized to $\pm 1$. In (\ref{spin1}) $|1u\rangle$ means that the spin component of particle 1 in some fixed ($z$) direction is $u$, while $|2v\rangle$ means that the spin component of particle 2 in the $z$-direction is $v$. This state expresses that the total spin of the two particles is 0. One can imagine that these particles previously have been together in some bound state with spin 0.

Alice is given the choice between measuring the spin component of her particle in one of two directions $a$ or $a'$. If she measures in the $a$-direction, her response ($\pm 1$) is called $A$, and if she measures in the $a'$ direction, her response is called $A'$. Similarly, Bob can measure in one of two directions $b$ (giving a response $B$) or $b'$ (giving a response $B'$). The whole procedure is repeated $n$ times with different entangled particle pairs and with different directions/ settings chosen by Alice and Bob.

We now temporarily take the point of departure that all these response variables exist in some sense. This can be seen as an assumption on realism. At the very least, we will assume that this point of departure is meaningful for some observers or for a group of communicating observers.

Since all responses then are $\pm 1$, we then have the inequality
\begin{equation}
AB+A'B+AB'-A'B' = A(B+B')+A'(B-B')\le 2.
\label{Bell2}
\end{equation}
The argument for this is simply: $B$ and $B'$ are either equal to one another or unequal. In the first case, $B-B'=0$ and $B+B'=\pm 2$; in the last case $B-B'=\pm 2$ and $B+B'=0$. Therefore, $AB+A'B+AB'-A'B'$ is equal to either $A$ or $A'$, both of these being $\pm 1$, multiplied by $\pm 2$. All possibilities lead to $AB+A'B+AB'-A'B' =\pm 2$.

From this, a statistician will argue: Assume that we can consider $A, A', B$ and $B'$ as random variables, defined on the same probability space $(\Omega, \mathcal{F}, P)$. Then by taking expectations over the terms in (\ref{Bell2}), we find
\begin{equation}
E(AB)+E(A'B)+E(AB')-E(A'B')\le 2.
\label{Bell3}
\end{equation}

A physicist will have a related argument: Assume that there is a hidden variable $\lambda$ such that $A=A(\lambda), A'=A'(\lambda), B=B(\lambda)$ and $B'=B'(\lambda)$. The assumption that such a hidden variable exists, is called local realism in the physical literature. By integrating over the probability distribution $\rho$ of $\lambda$, this gives
\begin{equation}
E(AB)=\int A(\lambda) B(\lambda) \rho (\lambda)d\lambda
\label{Bell4}
\end{equation}
etc.. Thus, by integrating term for term in (\ref{Bell2}), we again find (\ref{Bell3}).

 Of course, the above two arguments are equivalent; it is just a question of using either the notation $(\omega , P)$ or $(\lambda ,\rho )$. There are different traditions here. These arguments are reviewed and discussed in detail by Richard Gill (2014).
 
  The inequality (\ref{Bell3}) is called the CHSH inequality after the authors of Clauser et al. (1969), and has been the source of much controversy. First, it is known that if we use quantum mechanics to model the above experiment, one can find settings such that the CHSH inequality is violated. Secondly, recent loophole-free experiments (Giustina et al., 2015, Shalm et al., 2015, taken together with the experiments done by the 2022 Nobel laureates in physics) have shown that the CHSH inequality may be violated in practice. The CHSH inequality can be seen as a special, and important, Bell inequality.
  
  Thus, as a temporary conclusion, the simple assumptions sketched above for (\ref{Bell3}) cannot hold.
  
\section{The Hilbert space formulation from two different maximal accessible variables}

A completely new approach towards quantum foundations is proposed in Helland (2021, 2022a,b, 2023a,b). The basis is a model of an observer's mind when this observer is in some physical situation. In this situation there are physical variables, and the observer will have at least some of these variables, say $\theta, \lambda, \eta,...$ in his mind. Some of the variables are \emph{accessible} to him, that is, it is, in some future, in principle possible to obtain as accurate values as he wishes on the relevant variable. Others are \emph{inaccessible}. Examples of the latter are the vector (position, momentum) of a particle at some time, or the full spin vector of a spin particle.

The main assumption in the new model is as follows: \emph{Related to the mind of an observer $C$ there is an inaccessible variable $\phi$ such that all accessible variables can be seen as functions of $\phi$}. In the two examples above one can take $\phi =$ (theoretical position, theoretical momentum) and $\phi =$ full spin vector. In the last example, in the spin 1/2 case, one can model the discrete spin component in direction $a$ as $f^a(\phi)=\mathrm{sign}(\mathrm{cos}(a, \phi))$- Letting $\phi$ have a uniform distribution on the sphere gives the correct distribution of each $\theta^a$. Here these variables are physical variables, but following Zwirn (2016, 2020), every description of reality must be seen relative to the mind of some observer. Hence we can assume that the variables also exist in the mind of $C$.

Seen from a general point of view, the main assumption as formulated above is a rather strong assumption. But it has important consequences: From this and some rather weak assumptions the whole Hilbert space formalism can be deduced (Helland, 2023b). It is important that in this approach all variables, including $\phi$, can be seen just as \emph{mathematical} variables. 

In the two examples above, there are also maximal accessible variables: In the first example either position or momentum, in the second example the spin component $\theta^a$ in some direction $a$. In general, an accessible variable $\theta$ is maximal for an observer at time $t$ if there is no other accessible variable $\lambda$ at this time for the observer such that $\theta = f(\lambda)$ for some non-invertible function $f$.

Two different accessible variables $\theta$ and $\eta$ are said to be related if there is a transformation $k$ in $\phi$-space and a function $f$ such that $\theta = f(\phi)$ and $\eta = f(k\phi)$. In Helland (2023b) it is shown that if both $\theta$ and $\eta$ are maximal and take $r$ values, then $\phi$ can be constructed in a natural way such that $\theta$ and $\eta$ are related with respect to this $\phi$. In particular, two spin components $\theta^a$ and $\theta^b$ are related with respect to the total spin vector $\phi$. More specifically, they are related with respect to the projection $P\phi$ of $\phi$ upon the plane spent by the vectors $a$ and $b$.

If no such transformation $k$ can be found for a given $\phi$, we say that $\theta$ and $\eta$ are non-related with respect to this $\phi$.

In the present paper, we will concentrate on accessible variables that take a finite number of values. Then the following is proved in the above papers: Given a situation with two related maximal accessible variables, there corresponds to every accessible variable a self-adjoint operator in some Hilbert space $\mathcal{H}$. Under some weak postulates it can be proved that the eigenvalues of this operator are the possible values of the variable, say $\theta$, and that the eigenspaces are in one-to-one correspondence with questions `What is $\theta$?' together with sharp answers `$\theta=u$'. A variable is maximal if and only if all eigenspaces of the corresponding operator are one-dimensional.

The proof of this result relies on constructing a group $N$ acting upon $\psi = (\theta, \eta)$, where $\theta$ and $\eta$ are assumed to be related maximal accessible variables, and a certain representation $W(\cdot)$ of this group which is proved to be irreducible.

This is the first step in a new proposed foundation of quantum theory. For the present article, we do not need much more than this.

But one can also show the following (Helland, 2023b): If $k$ is the transformation connecting two related variables $\theta$ and $\eta$, then there is a unitary transformation $W(k)$ such that

 \begin{equation}
  A^\eta = W(k)^\dagger A^\theta W(k).
  \label{related}
  \end{equation}

On the other hand, if (\ref{related}) hold between two operators, then the corresponding variables $\theta$ and $\eta$ are related. This is proved by the following lemma.
\bigskip

\textbf{Lemma} \textit{Consider two maximal accessible finite-dimensional theoretical variables $\theta$ and $\lambda$. If there is a transformation $s$ of $\Omega_\psi$ such that $A^\lambda = W(s)^\dagger A^\theta W(s)$, then $\theta$ and $\lambda$ are related.}
\bigskip

\textit{Proof}
By equation (26) in Helland (2023b) the relation $A^\lambda = W(s)^\dagger A^\theta W(s)$ implies that
\begin{equation}
A^\lambda = \sum_n f_\theta (sn)|v_n\rangle\langle v_n|
\label{u8}
\end{equation}
for an orthonormal basis$\{|v_n\rangle\}$ for which we also have
\begin{equation}
A^\theta = \sum_n f_\theta (n)|v_n\rangle\langle v_n|.
\label{u9}
\end{equation}
But by the spectral theorem, we also have
\begin{equation}
A^\lambda = \sum_n f_\lambda (n)|v'_n\rangle\langle v'_n|
\label{u10}
\end{equation}
for some orthonormal basis. Since all the accessible variables are maximal, the eigenspaces are one-dimensional by Theorem A2 of Helland (2023b). In particular, this means that since $A^\lambda$ has a spectral decomposition in one basis in (\ref{u8}), the same basis must be used in (\ref{u10}). This also implies
\begin{equation}
\lambda(n) = f_\lambda (n)=f_\theta(sn)=\theta(sn),
\label{u11}
\end{equation}
so $\theta$ and $\lambda$ are related with respect to the elements of the group $N$. Since by the model, each variable is a function of the inaccessible variable $\phi$, they must also be related with respect to $\phi$. 
$\Box$
\bigskip

Given the results above, a rich theory follows. The set of eigenvalues of the operator $A^\theta$ is identical to the set of possible values of $\theta$. The variable $\theta$ is maximal if and only if all eigenvalues are simple. In general, the eigenspaces of $A^\theta$ are in one-to-one correspondence with questions `What is $\theta$'/ `What will $\theta$ be if we measure it?' together with sharp answers $\theta=u$.

This points at a new foundation of quantum theory, and it also suggests a general epistemic interpretation of the theory: Quantum theory is not directly a theory about the world, but a theory about an actor's knowledge of the world. From this conclusion also a number of so-called `quantum paradoxes' can be solved (assuming a version of quantum theory in which the only acceptable pure states are those which are eigenvectors of some physically meaningful operator), see Helland (2022b, 2023a,c).
  
 \section{Limitation in the mind of an observer in some context}
 
 Let an observer $O$ be in some physical context. In agreement with his observations and planned observations, he will have several physical variables $\theta, \eta,\lambda,...$ in his mind. I will assume here that all these variables are finite-dimensional. According to my general model, these variables can all be seen as functions of a fixed inaccessible variable $\phi$. Then we have the following:
 \bigskip
 
 \textbf{Theorem 1} \textit{Assume that $O$ has two related maximal accessible variables $\theta$ and $\eta$ in his mind at some fixed time $t$.  Then he cannot simultaneously have in his mind another maximal accessible variable that is related to $\theta$, but non-related to $\eta$.}
 \bigskip
 
 \underline{Proof}
 
 Denote the new variable by $\xi$. Let the transformation relating $\theta$ and $\eta$ be $k$, and the transformation relating $\theta$ and $\xi$ be $s$. Then by (\ref{related}) we have $A^\theta = W(k) A^\eta W(k)^\dagger$ and $A^\theta = W(s) A^\xi W(s)^\dagger$. But this implies that $A^\eta = W(k^{-1}s) A^\xi W(k^{-1}s)^\dagger$, so we conclude from the Lemma of the previous Section that $\eta$ and $\xi$ are related, which leads to a contradiction with the assumptions made.
 $\Box$
 \bigskip
 
 \textit{Note:} It is essentially here that a fixed context is assumed, and that the time $t$ is kept fixed. If we allow time to vary, $O$ can have in mind a large number of variables, also non-related ones. Also, note that $O$ here can be any person.
 
  \section{Limitations in the mind of Charlie}
  
  Assume that Alice and Bob have made $n$ independent repetitions of the Bell experiment, each time choosing their settings either by their own free will or by some random mechanism. After this, they leave all their data to a statistician Charlie, who will try to make a statistical model of the experiment.
  
  To this end, he thinks of a typical experiment, with setting either $a$ or $a'$ chosen by Alice, giving corresponding responses $A$ or $A'$, each either -1 or +1, and similarly settings $b$ or $b'$ chosen by Bob, with responses $B$ and $B'$. As a good statistician, he relies on the conditionality principle (see for instance Cox, 1958), and wants to condition his model upon all the settings, only modelling the possible responses.
  
The inaccessible variables, in this case, are the full spin vector $\phi_A$ of Alice's particle and the full spin vector $\phi_B$ of Bob's particle. To Charlie, both these are inaccessible, but at the same time, he knows from other experiments that if Alice and Bob measure spin in the same direction, then their spin components are opposite.
Thus, in his mind, he may think of a model where $\phi_B = - \phi_A$, in the sense that any measured components are opposite. For a precise statement and proof based upon the entangled state vector (\ref{spin1}), see Section 2.

The goal of Charlie is to formulate a model for the possible responses  $A, A', B$ and $B'$. Each of these variables takes the values -1 or +1, and each is maximal as a variable that is accessible to respectively Alice and Bob. For Charlie, we will see that the variables are pairwise related under the model. During his modelling attempts, he is able to communicate with Alice and Bob. The term `accessible' will be defined with respect to this whole set of communicating actors, meaning that a variable is called accessible if it at the outset also has been accessible either to Alice or to Bob. In this sense, each of the variables  $A, A', B$ and $B'$ must be seen as maximal accessible.
\bigskip

\textbf{Theorem 2} \textit{Under these assumptions, in relation to Charlie's model of the variables $A, A', B$ and $B'$, the following hold: 1) All the pairs of variables are related under his model. 2) Given two such pairs which share one variable, the fourth variable will be related to the variables of both pairs, but non-related to another variable, a function of the variables in the two pairs.}
\newpage

\underline{Proof}

1) Look first at a pair associated with the same primitive actor, say, Alice, so the variables are $A$ and $A'$. These are spin components in different directions, and Charlie can have, as an inaccessible part of his model, the projection $\lambda$ of Alice's spin vector upon the plane spent by these two directions. Let $k$ be a suitable rotation of $\lambda$ in this plane. Then for some value of $\lambda$ and for some function $f$ we have $A=f(\lambda)$ and $A' =f(k\lambda)$. Explicitly, the function $f$ can be taken as $f(\lambda)= \mathrm{sign}(\mathrm{cos}(a,\lambda))$.

Then look upon a pair of variables from both primitive actors, which possibly can be measured in the same run, say $A$ and $B$. Then by what has been proved above, $A$ is related to a hypothetical variable $A''$, measured by Alice in the Bob-direction $b$, which is known to her. But, by Charlie's mental model, he must conclude that $A''=-B$, hence $A$ and $B$ are related.

2) Say that the two pairs are $(A, B)$ and $(A, B')$, so that the fourth variable is $A'$. Then $A'$ is related to both variables in the first pair, and to both variables in the second pair by 1) above. Now look at the variable $C=A(B+B')$. This variable takes the value $0$ if $B=-B'$, and the values $\pm 2$ if $B=B'$. This variable is determined by the joint values of $A, B$ and $B'$, but is non-related to the value of $A'$. Thus, we must conclude that $C$ and $A'$ are non-related; they cannot be coupled by any transformation in Charlie's $\phi$-space. The same is true for $D=|C|-1$ and $A'$. Note that $D$ takes the values $\pm 1$, and thus is on the same scale as $A'$.
$\Box$
\bigskip

Thus, we are forced to conclude from Theorem 1 and Theorem 2:
\bigskip

\textbf{Corollary 1} \textit{Charlie is not able to keep all four variables $A, A', B$ and $B'$ in his mind at the same time $t$ during the modelling process.}
\bigskip

Now go back to the discussion of Section 3, and the assumptions made there for deriving the inequality (\ref{Bell2}):

`We temporarily take the point of departure that all these response variables
exist in some sense. This can be seen as an assumption on realism. At the very least,
we will assume that this point of departure is meaningful for some observers or for a
group of communicating observers.'

By Corollary 1, this assumption does not hold for the observer Charlie. In particular then, the assumptions behind the derivation there of the CHSH inequality (\ref{Bell3}) do not hold. This explains why this inequality can be violated in practice.

Note that Charlie may be any observer, and he may be able to communicate with any person. This person may be interested in the Bell experiment, and he may after the experiment have had contact with Alice and Bob, and have listened to their experiences.
\bigskip

\textbf{Corollary 2} \textit{For any communicating person there may exist situations where he is not able to keep in his mind all the variables that he is exposed to.}
\bigskip

In particular, this seems to happen for any person trying in practice in a concrete context to reproduce the simple argument leading to (\ref{Bell3}) in Section 3 above in a way that can be communicated to all people, including the actors Alice and Bob at a time when they make Bell experiments, taking into account \emph{their} limitations at such times.

\section{Bell's theorem}

There are a few variants of Bell's theorem (Bell, 1964) in the literature, and the theorem has led to several rather complicated discussions among theoretical physicists. According to Wikipedia, the theorem has two components: First, physical variables exist independently of being observed or measured (sometimes called the assumption of realism); and second, Alice's choice of action cannot influence Bob's result or vice versa (often called the assumption of locality).

Bell's theorem then says that these two assumptions imply that the Bell inequality/ CHSH inequality hold whenever a Bell experiment is performed. So when we now know for sure that this inequality may be violated in practice, one of the assumptions must be abandoned.

In other discussions, one agrees about the Bell theorem as a mathematical theorem, and the violation of the CHSH inequality in practice is claimed to require either non-locality or lack of statistical independence. Statistical independence must be formulated with respect to some mathematical model, a model which again assumes some form of realism.

From the treatment given in this article, it seems clear that it is the assumption of realism that must be given up. My own point of departure is similar to Herv\'{e} Zwirn's Convivial Solipsism (Zwirn, 2016, 2020): Every description of reality must be relative to the mind of an observer (or relative to the joint minds of a communicating group of observers). To be in agreement with observations, we seem to be forced to the conclusion that the minds of these observers must be limited in certain situations.

 \section{Some conclusions}
 
 Let us try to recapitulate. The Bell experiment has to do with two actors Alice and Bob who cannot communicate during the experiment. They both make dichotomous decisions on their settings of some apparatus, and they both observe dichotomous responses. For any actor who tries to understand these responses, according to my theory, his mind must be limited: He is simply not able to keep enough variables in his mind at the same time.
 
 If this is true, it also must have consequences for other situations. It is well known that quantum theory may be used to model the process of making decisions, see for instance Busemeyer and Buza (2012). A new foundation of quantum decision theory, based on the theory of Helland (2023b), is proposed in Helland (2023e).
 
 So assume in general that Alice and Bob belong to different groups of people that do not communicate, and they both make a series of decisions, say dichotomous ones, and these decisions have consequences, which are accessible to Alice and Bob, respectively, in the situation, and in a certain sense maximally so. Later, the consequences may be accessible to outsiders, but by arguments given in the present article, these outsiders may sometimes have difficulties in analyzing the consequences in a complete and satisfactory way. They may simply not be able to keep enough variables in their minds at the same time.
 
 Such general descriptions may seem to be relevant to many decision situations, but to go into a detailed discussion of such stuations, will involve considerations that go far beyond physics.

The main postulate of Helland (2023b), that all accessible variables may be seen as functions of an inaccessible variable $\phi$, may well be seen as an assumption of superdeterminism. Religious aspects of this postulate (Helland, 2023f) are now being discussed in a Dialogo conference, a conference promoting dialogues between scientists and theologians.

\section*{References}

$\ \ \ \ \ \ $Bell, J.S. (1964). On the Einstein-Podolsky-Rosen paradox. Physics 1, 195-200.

Busemeyer, J.R. and Buza, P.D. (2012). Quantum Models for Cognition and Decisions. Cambridge University Press, Cambridge.

Clauser, J.F., Horne, M.A., Shimony, A. and Holt, R.A. (1969). Proposed experiment to test local hidden-variables theories. Phys. Rev. Lett. 23, 880-884.

Cox, D. (1958). Some problems connected with statistical inference. Annals of Statistics 29, 357-372.

Gill, R.D. (2014). Statistics, causality, and Bell's theorem. Statistical Science 20 (4), 512-528.

Giustina, M., Versteegh, M.A.M, Wengerowsky, S. et al. (2015). Significant-loophole-free test of Bell's theorem with entangled photons. Phys. Rev. Lett. 115, 250401.

Helland, I.S. (2019).  When is a set of questions to nature together with sharp answers to those questions in one-to-one correspondence with a set of quantum states? arXiv: 1909.08834 [quant-ph].

Helland, I.S. (2021). Epistemic Processes. A Basis for Statistics and Quantum Theory. Springer, Cham, Switzerland.

Helland, I.S. (2022a). The Bell experiment and the limitation of actors. Foundations of Physics 52, 55.

Helland, I.S. (2022b). On reconstructing parts of quantum theory from two related maximal conceptual variables. Int. J. Theor. Phys. 61, 69.

Helland, I.S. (2023a). On the Foundation of Quantum Theory. The Relevant Articles. Eliva Press. Chisinau, Moldova.

Helland, I.S. (2023b). An alternative foundation of quantum theory. arXiv: 2305.06727 [quant-ph]. To appear in Foundations of Physics.

Helland, I.S. (2023c). Possible connections between relativity theory and a version of quantum mechanics based on theoretical variables. arXiv: 2305.15435 [physics-hist.ph].

Helland, I.S. (2023d). On the Bell experiment and quantum foundation. Journal of Modern and Applied Physics 6 (2), 1-5.

Helland, I.S. (2023e). On the foundation of quantum decision theory. Preprint.

Helland, I.S. (2023f). Quantum mechanics as a theory that is consistent with the existence of God. Discussion paper to a Dialogo conference.

H\"{o}hn, P.A. and Wever, C.S.P. (2017). Quantum theory from questions. Physics Reviews A, 95, 012102.

Shalm, L.K., Meyer-Scott, E. Christensen, B.G. Bierhorst, P. et al. (2015). Strong loophole-free test of local realism. Phys. Rev. Lett. 115, 250402.

Susskind, L. and Friedman, A. (2014). Quantum Mechanics. The theoretical Minimum. Basic Books, New York.

Zwirn, H. (2016). The measurement problem: Decoherence and convivial solipsism. Foundations of Physics 46, 635-667.

 Zwirn, H. (2020). Nonlocality versus modified realism. Foundations of Physics 50, 1-26.

\end{document}